\documentclass[twocolumn]{webofc}

\usepackage[varg]{txfonts}   % Web of Conferences font
\usepackage{hyperref}
\usepackage{url}
\hypersetup{colorlinks=true,citecolor=blue,urlcolor=blue,linkcolor=blue}
\usepackage{upgreek}
\usepackage{newtxmath}
\usepackage{siunitx}
\DeclareRobustCommand{\upe}{\ifmmode{\mathrm{e}}\else{e}\fi}
\DeclareRobustCommand{\muegamma}{\ifmmode{\upmu\to \upe\upgamma}\else{\(\upmu\to \upe\upgamma\)}\fi}
\DeclareRobustCommand{\muegammasign}{\ifmmode{\upmu^+\to \upe^+\upgamma}\else{\(\upmu^+\to \upe^+\upgamma\)}\fi}
\DeclareRobustCommand{\rmd}{\ifmmode{\upmu\to \upe\upnu\upnu\upgamma}\else{\(\upmu\to \upe\upnu\upnu\upgamma\)}\fi}
\DeclareRobustCommand{\michel}{\ifmmode{\upmu\to \upe\upnu\upnu}\else{\(\upmu\to \upe\upnu\upnu\)}\fi}
\DeclareRobustCommand{\photon}{\ifmmode{\upgamma}\else{\(\upgamma\)}\fi}
\DeclareRobustCommand{\egamma}{\ifmmode{E_{\upgamma}}\else{\(E_{\upgamma}\)}\fi}
\DeclareRobustCommand{\tgamma}{\ifmmode{t_{\upgamma}}\else{\(t_{\upgamma}\)}\fi}
\DeclareRobustCommand{\xgamma}{\ifmmode{\vec{x}_{\upgamma}}\else{\(\vec{x}_{\upgamma}\)}\fi}
\DeclareRobustCommand{\Nsum}{\ifmmode{N_{\mathrm{sum}}}\else{\(N_{\mathrm{sum}}\)}\fi}
\DeclareRobustCommand{\Nmppc}{\ifmmode{N_{\mathrm{MPPC}}}\else{\(N_{\mathrm{MPPC}}\)}\fi}
\DeclareRobustCommand{\Npmt}{\ifmmode{N_{\mathrm{PMT}}}\else{\(N_{\mathrm{PMT}}\)}\fi}
\DeclareRobustCommand{\Npho}{\ifmmode{N_{\mathrm{pho},i}}\else{\(N_{\mathrm{pho},i}\)}\fi}
\DeclareRobustCommand{\Nphe}{\ifmmode{N_{\mathrm{phe},i}}\else{\(N_{\mathrm{phe},i}\)}\fi}

\begin{document}
\title{Photon energy reconstruction with the MEG~II liquid xenon calorimeter}
%
% subtitle is optional
%
%%%\subtitle{Do you have a subtitle?\\ If so, write it here}

\author{\firstname{Kensuke} \lastname{Yamamoto}\inst{1}\fnsep\thanks{\email{kensukey@icepp.s.u-tokyo.ac.jp}} \and
        \firstname{Sei} \lastname{Ban}\inst{1} \and
        \firstname{Lukas} \lastname{Gerritzen}\inst{1} \and
        \firstname{Toshiyuki} \lastname{Iwamoto}\inst{1} \and
        \firstname{Satoru} \lastname{Koboyashi}\inst{1} \and
        \firstname{Ayaka} \lastname{Matsushita}\inst{1} \and
        \firstname{Toshinori} \lastname{Mori}\inst{1} \and
        \firstname{Rina} \lastname{Onda}\inst{1} \and
        \firstname{Wataru} \lastname{Ootani}\inst{1} \and
        \firstname{Atsushi} \lastname{Oya}\inst{1}
        % etc.
}

\institute{ICEPP, The University of Tokyo, 7-3-1 Hongo, Bunkyo-ku, Tokyo 113-0033, Japan
          %\and %the second here if necessary
}

\abstract{The MEG~II experiment searches for a charged-lepton-flavour-violating \muegamma{} with the target sensitivity of \num{6e-14}.
A liquid xenon calorimeter with VUV-sensitive photosensors measures photon position, timing, and energy.
This paper concentrates on the precise photon energy reconstruction with the MEG~II liquid xenon calorimeter.
Since a muon beam rate is \qtyrange[range-units=single,range-phrase=--]{3}{5e7}{\per\second}, multi-photon elimination analysis is performed using waveform analysis techniques such as a template waveform fit.
As a result, background events in the energy range of \qtyrange[range-units=single,range-phrase=--]{48}{58}{\MeV} were reduced by \qty{34}{\percent}.
The calibration of the energy scale of the calorimeter with several calibration sources is also discussed to achieve a high resolution of \qty{1.8}{\percent}.
}
\maketitle
%
%=================================================
\section{Introduction}
\label{sec:intro}

The charged-lepton-flavour-violating muon rare decay, \muegamma{}, is strongly suppressed below a branching ratio of \num{e-54} in the Standard Model of particle physics considering neutrino oscillation.
A branching ratio of \numrange[range-phrase=--]{e-11}{e-14} is, however, predicted by new physics models beyond the Standard Model, such as supersymmetry \cite{Kuno-okada}.
\begin{figure}[bp]
  \centering
  \includegraphics[width=8cm,clip]{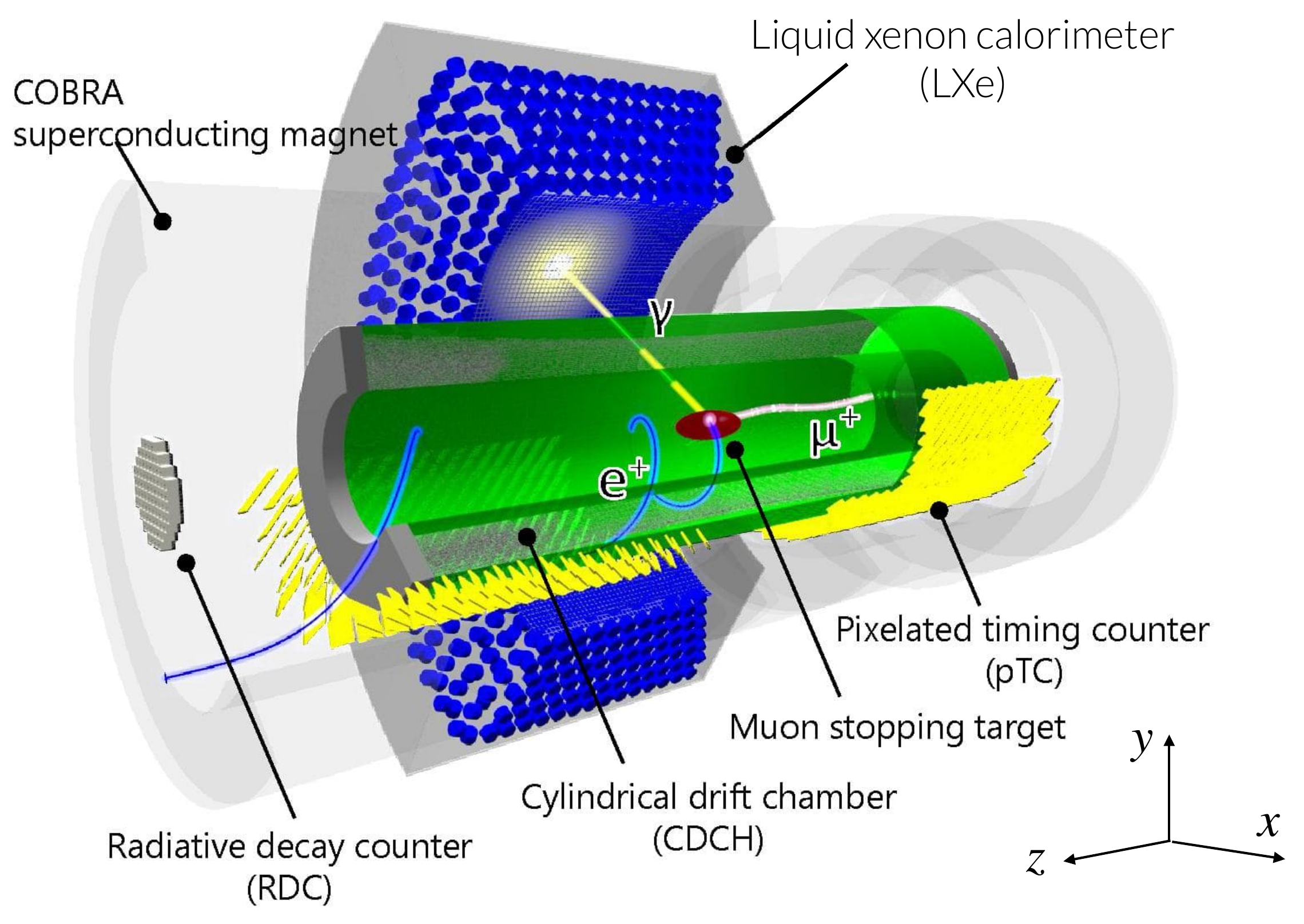}
  \caption{A sketch of the MEG~II detector. Partially modified from Ref.~\cite{meg2-design}.}
  \label{fig:meg2-detector}
  \vspace{-1\baselineskip}
\end{figure}
The MEG~II experiment aims to search for \muegamma{} with a sensitivity of approximately \num{6e-14} using innovative high-resolution detectors and the most intense DC muon beam at Paul Scherrer Institut \cite{meg2-design, meg2-detector} (Figure~\ref{fig:meg2-detector}).
%It started physics data-taking in 2021 and is planned to be continued to 2026.
Physics data-taking began in 2021 and is expected to continue until 2026.
An upper limit on the branching ratio was set to \num{3.1e-13} (\qty{90}{\percent} C.L.) with a combination of the MEG full dataset and the MEG~II first dataset \cite{meg2-2021}.

%Precise measurements of photons and positions realise high sensitivity with the kinematical difference between signal and background.
Precise positron and photon measurements achieve high sensitivity by exploiting the kinematic difference between signal and background.
The signal kinematics is a two-body decay: positron and photon are emitted back-to-back with monochromatic energy of \qty{52.8}{MeV} simultaneously.
Meanwhile, the dominant background is an accidental coincidence of positron and photon from different parent muons.
The positron background source is the Michel decay (\michel{}), and the photon background is generated by radiative muon decay (RMD; \rmd{}) and positron annihilation in flight (AIF) with electrons in the positron spectrometer.
The number of the accidental background $N_{\mathrm{acc}}$ depends on the detector resolution \(\sigma\):
\begin{equation}
  N_{\mathrm{acc}} \propto \sigma_{\egamma{}}^2 \cdot \sigma_{E_{\upe}} \cdot \sigma_{\Theta_{\upe\upgamma}}^2 \cdot \sigma_{t_{\upe\upgamma}},
  \label{eq:acc-bg}
\end{equation}
where \(E_{\upgamma(\upe)}\) is the photon (positron) energy, \(\Theta_{\upe\upgamma}\) is the opening angle between positron and photon, and \(t_{\upe\upgamma}\) is the time difference between them.

This paper will concentrate on the photon energy measurement for the data taken in 2021 and 2022.
A liquid xenon calorimeter as a photon detector is introduced in Section~\ref{sec:xec}.
We then discuss analysis methods to realise high-resolution energy measurements in a high-intensity muon beam.
Section~\ref{sec:pileup} presents a multi-photon elimination algorithm.
Section~\ref{sec:egamma} discusses the calibration of the energy scale of the calorimeter.

%=================================================
\section{Liquid xenon calorimeter}
\label{sec:xec}

\begin{figure}[tbp]
  \centering
  \includegraphics[width=8cm,clip]{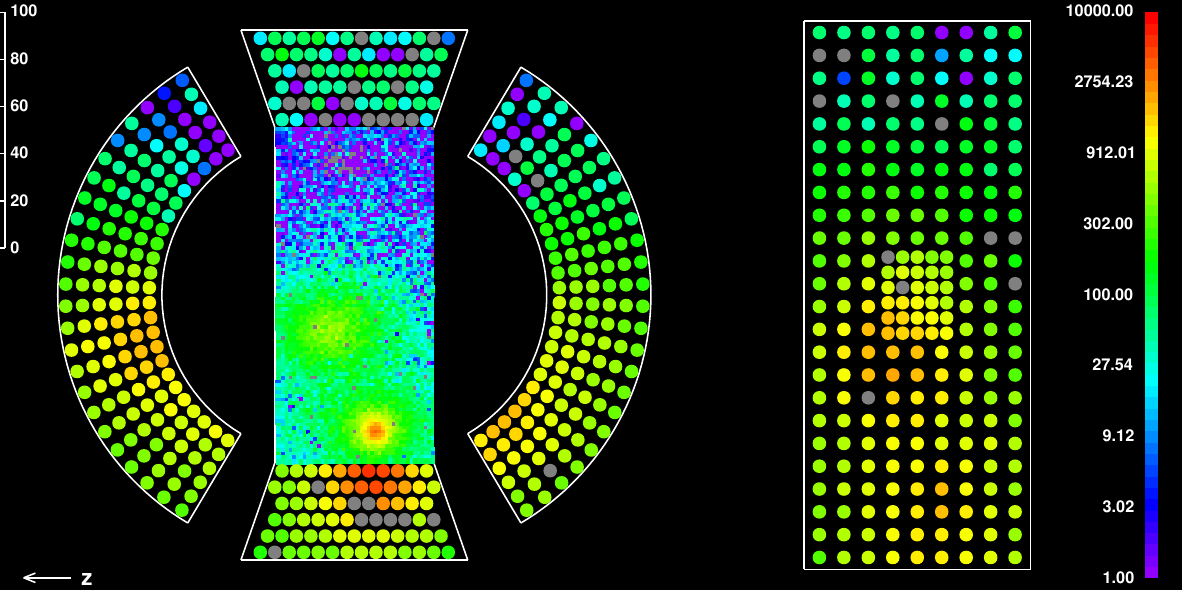}
  \caption{An unfolded view of the LXe calorimeter and scintillation light distribution. A colour bar represents the number of scintillation photons \Npho{} on the \(i\)-th photosensor. Two \photon{}-rays impinge on the bottom half of the calorimeter.}
  \label{fig:event-display}
  \vspace{-1\baselineskip}
\end{figure}

A liquid xenon (LXe) calorimeter plays a role in measuring photon position, timing, and energy in the MEG~II experiment.
This is a C-shape homogeneous calorimeter as shown in Figure~\ref{fig:meg2-detector} filled with \qty{900}{\liter} of LXe to obtain a uniform response.
The LXe has a lot of advantages in detecting \qty{52.8}{\MeV} photons; e.g. high stopping power (radiation length of \qty{2.8}{\cm}) and fast response (\qty{45}{\ns} decay time).
Since the scintillation light emitted from xenon is in the vacuum ultraviolet (VUV) region, \num{4760} VUV-sensitive photosensors are utilised for scintillation light detection.
The photon entrance face is covered by \num{4092} Multi-Pixel Photon Counters (MPPCs) with the size of \qtyproduct[product-units=single]{1.5 x 1.5}{\cm\squared} \cite{vuv-mppc} to achieve high granularity.
The other faces are covered by \num{668} 2-inch round-shape photomultiplier tubes (PMTs).
%The gain, photon detection efficiency of MPPCs, and quantum efficiency of PMTs
The photosensors were calibrated to reconstruct the number of impinging scintillation photons on the $i$-th photosensor \Npho{} (Figure~\ref{fig:event-display}) as presented in Section~6.3 of Ref.~\cite{meg2-detector}.
The signal waveforms are read out with the DRS4 waveform digitiser \cite{drs4} with a sampling frequency of \qty{1.4}{GSPS}.
%The local coordinates $(u,v,w)$ are defined as
%
%\begin{equation}
%  \begin{split}
%    u &= z, \\
%    v &= \arctan{\left(-\frac{y}{x}\right)} \cdot R_{\mathrm{in}}, \\
%    w &= \sqrt{x^2 + y^2} - R_{\mathrm{in}},
%  \end{split}
%  \label{eq:local-coordinate}
%\end{equation}
%
%where $R_{\mathrm{in}}$ is the inner radius of the active volume of the LXe calorimeter which is \qty{64.76}{cm}.

The photon energy is reconstructed by collecting all scintillation photons and converting them:
\begin{alignat}{2}
  %\egamma &= \Nsum \cdot S \cdot T(t) \cdot U(u_{\upgamma}, v_{\upgamma}, w_{\upgamma}) \label{eq:egamma} \\
  \egamma &= \Nsum \cdot S \cdot T(t) \cdot U(\xgamma), \label{eq:egamma} \\
  \Nsum &= \Nmppc \cdot r_{\mathrm{MPPC}}(t) + \Npmt, \label{eq:nsum}
\end{alignat}
where \(S\) is an energy scale conversion factor; \(T(t)\) and \(r_{\mathrm{MPPC}}(t)\) are temporal variation correction functions;
%\(U(u_{\upgamma}, v_{\upgamma}, w_{\upgamma})$\)
\(U(\xgamma)\) is a non-uniformity correction function; and \(N_{\mathrm{MPPC(PMT)}}\) is the weighted sum of the number of scintillation photons \Npho{} detected by MPPCs (PMTs).
%The weighted sums \Nmppc{} and \Npmt{} are calculated for a single photon through a multi-photon elimination algorithm discussed in Section~\ref{sec:pileup}.
The weighted \Nmppc{} and \Npmt{} sums are calculated, for single-photon events, via the multi-photon elimination algorithm discussed in Section~\ref{sec:pileup}.
%The conversion factor to the energy $S$, and correction functions $T(t)$, $r_{\mathrm{MPPC}}(t)$, and $U(\xgamma)$ are necessary to be calibrated to achieve high resolution as discussed in Section~\ref{sec:egamma}.
To achieve high resolution, the energy conversion factor \(S\) and the correction functions \(T(t)\), \(r_{\mathrm{MPPC}}(t)\), and \(U(\xgamma)\) must be calibrated as discussed in Section~\ref{sec:egamma}.

%=================================================
\section{Multi-photon elimination}
\label{sec:pileup}

Multi-photon events deteriorate the energy resolution in a high-intensity muon beam.
%Multiple \photon{}-rays impinge on the calorimeter in the DRS time window of approximately \qty{700}{ns} physically and accidentally in several events.
One of the multi-photon sources is two photons from AIF, both incident on the calorimeter.
%This type of event has to be discarded from the analysis sample because signal photons are unlikely to be detected with other coincident photons due to the event signature.
%this is only the coincident photon source in the MEG~II experiment.
%Meanwhile, not only radiative decay and AIF photons but also signal photons can be detected with accidental low-energy photons in the DRS time window of approximately \qty{700}{ns}, so-called ``pileup'', since beam muons stop at high rates of $3\text{--}\qty{5e7}{\per\second}$.
Furthermore, due to the high rate of the muon beam (\qtyrange[range-phrase=--,range-units=single]{3}{5e7}{\per\second}), not only background but signal photons can be detected in accidental coincidence with low-energy photons in the DRS time window of about \qty{700}{\ns}, the so-called "pileup".
The pileup event has to be unfolded to extract the information of each individual photon.

We first perform a peak search in the spatial distribution shown in Figure~\ref{fig:event-display} to identify the multi-photon event candidates.
We then analyse the MPPC and PMT summed waveforms with a template waveform fitting technique to determine whether the detected multiple photons are coincident and to unfold multiple pulses in case the multiple photons are out of time.
The template waveforms \(f(\tau)\) were created by taking an average of the measured summed waveforms.
Their fluctuation is expressed as a standard deviation at time \(\tau\) \(\sigma_f(\tau)\).
The template waveform fit minimises a \(\chi^2\) defined as
\begin{equation}
  \chi^2 = \sum_{\mathrm{MPPC,PMT}} \int \frac{\left( V(\tau) - \sum_{i}^{n_{\mathrm{pulse}}} f(\tau; A_i, t_i) \right)^2}{\sigma_f^2(\tau)} d\tau,
\end{equation}
where \(V(\tau)\) is observed MPPCs or PMTs summed waveform with position-dependent weights to optimise the \Nsum{} resolution, and \(n_{\mathrm{pulse}}\) is the number of fitted pulses.
The amplitude \(A_i\) and timing \(t_i\) for the \(i\)-th pulse are fit parameters.
Initial parameter sets of \(n_{\mathrm{pulse}}\), \(A_i\), and \(t_i\) are calculated with three techniques, discussed in Section~\ref{subsec:pileup-initial}, before the waveform fit is performed.
%These multi-photon detection techniques are important to make the fitting robust and improve the energy resolution.
%The performance evaluation of the multi-photon elimination is presented in Section~\ref{subsec:pileup-performance}.

%============================================
\subsection{Multi-photon detection techniques}
\label{subsec:pileup-initial}

\begin{figure}[tbp]
  \centering
  \includegraphics[width=8cm,clip]{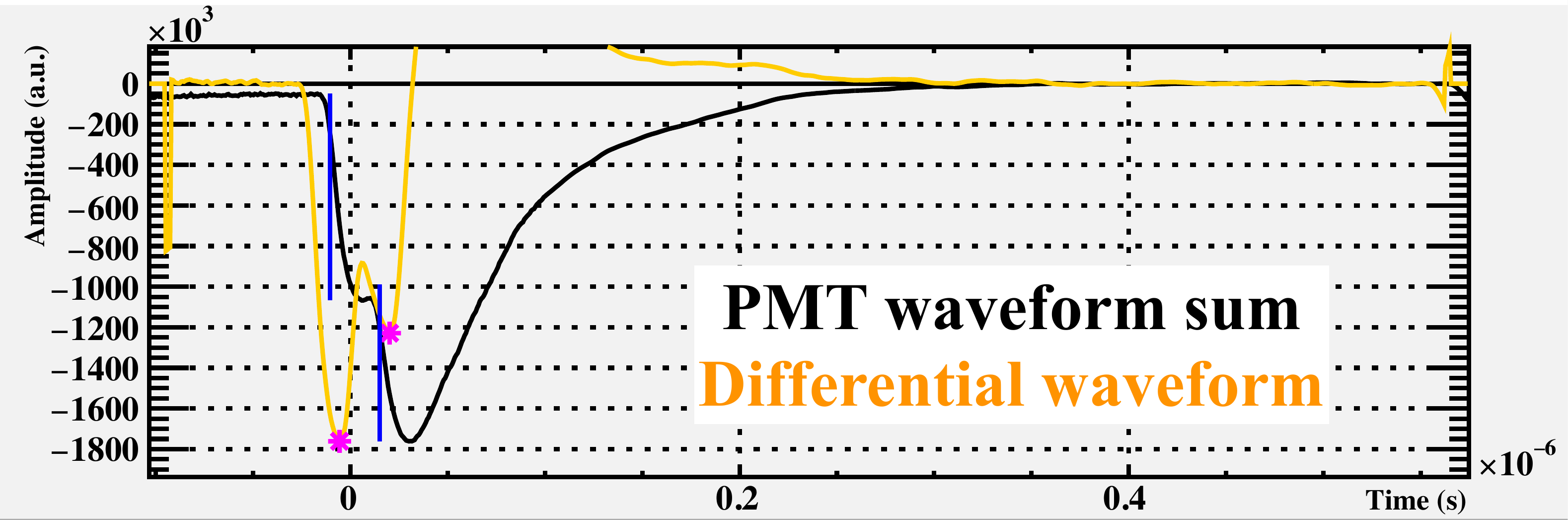}
  \caption{Multi-peak search in the PMT differential waveform in the same event as Figure~\ref{fig:event-display}. Magenta markers show the detected peaks in the differential waveform. Blue lines show the calculated peak time.}
  \label{fig:pileup-diff}
  \vspace{-1\baselineskip}
\end{figure}

The first step is to find pileup photons in the time domain from the summed waveform of photosensors.
Here, we search for pileups from the PMT waveform, which is sharper than MPPC's.
To make the waveform even sharper, we apply an additional algorithmic waveform processing to create a differential waveform dedicated to the pileup search.
As shown in Figure~\ref{fig:pileup-diff}, this processing makes waveform peaks (magenta markers) more distinguishable than those before the waveform processing.
It, therefore, enables efficient detection of pileups.
%
%The first step is the peak search in the differential PMT-summed waveform.
%We differentiate the PMT summed waveform because the waveform is sharper than that of MPPCs.
%Figure~\ref{fig:pileup-diff} shows PMT summed waveform and differential waveform.
%The differential one has two peaks shown by the magenta markers in Figure~\ref{fig:pileup-diff}, but not clear to see in the summed one.
This technique can distinguish pulses with \qtyrange[range-phrase=--,range-units=single]{15}{20}{\ns} and larger time differences according to the amplitude of the pulses.
%It is clear that the differential one is sensitive to multiple pulses.

\begin{figure*}[t]
  \centering
  %\vspace*{1cm} % Give the correct figure height in cm
  \includegraphics[width=14cm,clip]{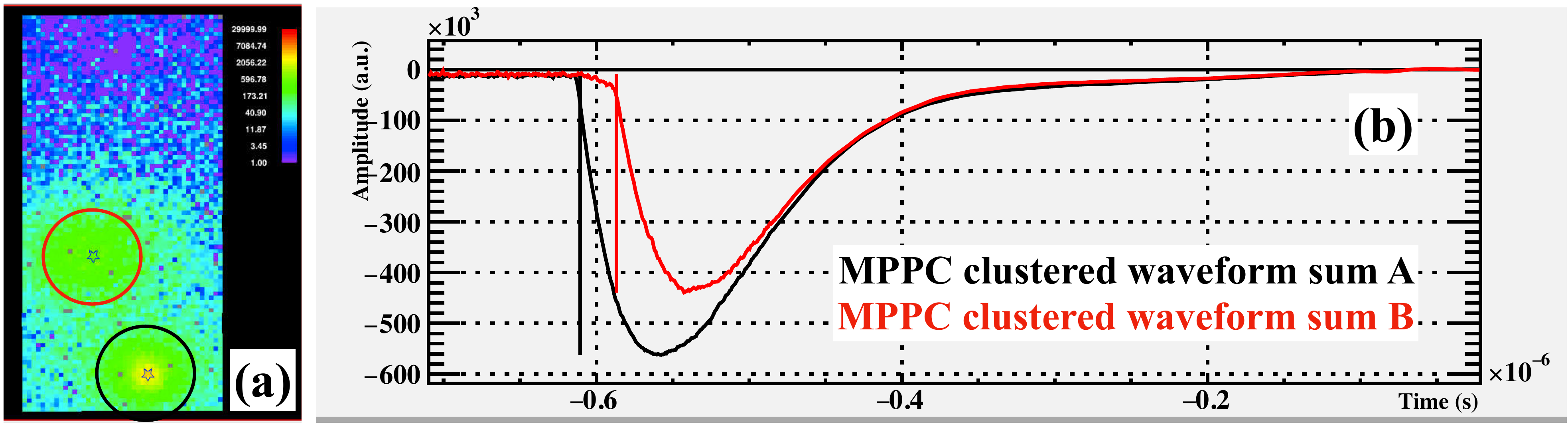}
  \caption{Multi-peak search based on the spatial distribution for the event of Figure~\ref{fig:event-display}.
           (a) Spatial distribution.
           (b) Summed waveforms of each cluster. The lines show the estimated pulse time.}
  \label{fig:pileup-clustering}
  \vspace*{-1\baselineskip}
\end{figure*}

The second step is clustering the photosensors based on the spatial distribution and analysing the summed waveforms of each cluster shown in Figure~\ref{fig:pileup-clustering}.
This technique can distinguish distant photons with small time differences of a few to \qty{20}{\ns}, which are not distinguished by the differential waveform.

%\begin{figure}[tbp]
%  \centering
%  \includegraphics[width=8cm,clip]{pileup-fadc.pdf}
%  \caption{The flash ADC signal used for the trigger. Pulse search before the DRS time window takes advantage of the wide time window of \qty{1600}{ns}. Blue lines show the calculated pulse timings. The event differs from that shown in Figures~\ref{fig:event-display}, \ref{fig:pileup-diff}, \ref{fig:pileup-clustering}.}
%  \label{fig:pileup-fadc}
%\end{figure}

The last step is using the flash ADC signal used for the trigger whose time window is \qty{1600}{\ns}.
%(Figure~\ref{fig:pileup-fadc}).
The time window is more than twice as wide as the DRS time window (approximately \qty{700}{\ns}), though its sampling frequency is \qty{80}{MSPS}.
This signal enables us to obtain information on pulses coming before the DRS time window.

\begin{figure}[tbp]
  \centering
  \includegraphics[width=8cm,clip]{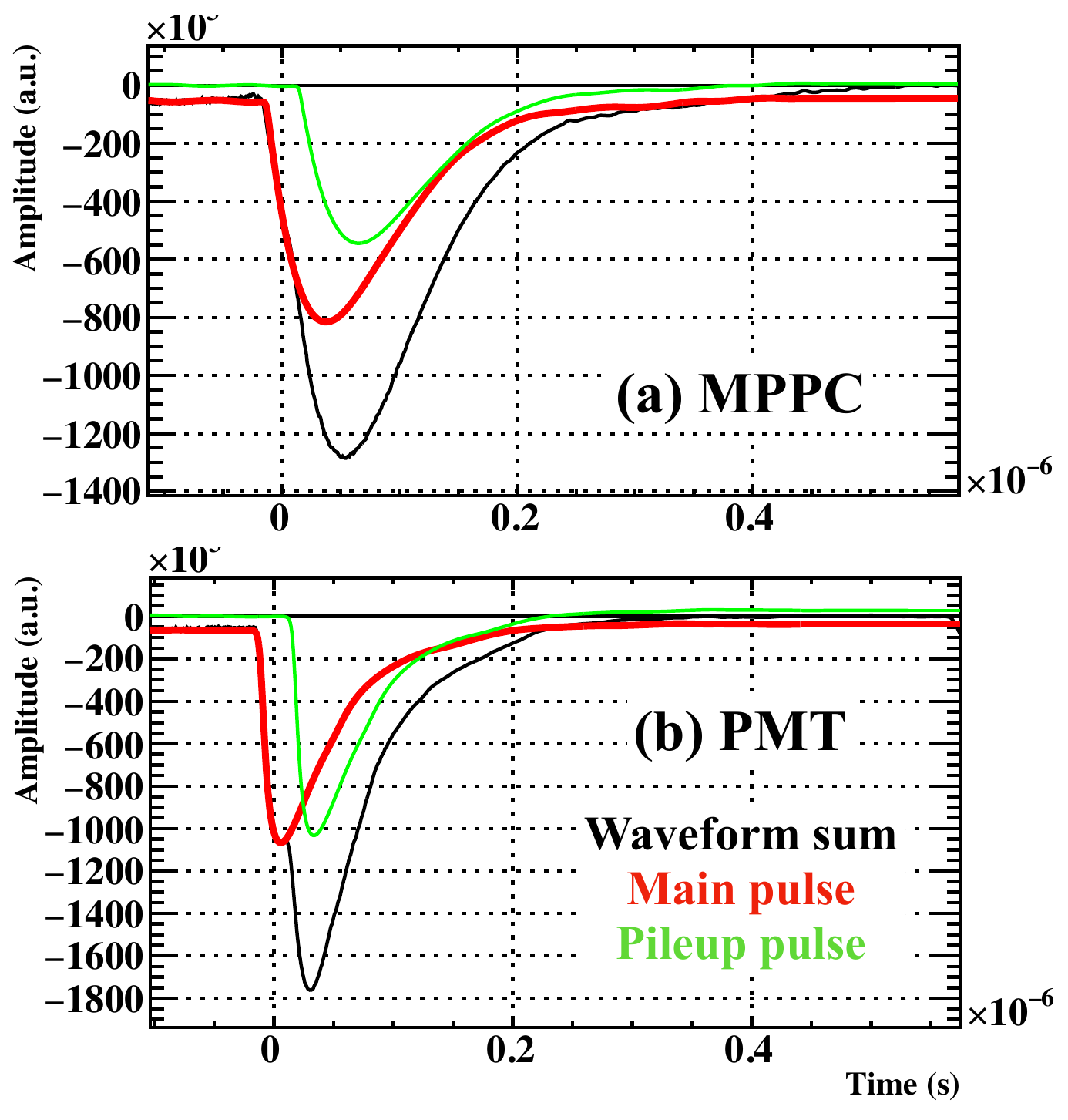}
  \caption{Unfolded multiple pulses for the event shown in Figures~\ref{fig:event-display}, \ref{fig:pileup-diff}, \ref{fig:pileup-clustering}. The weighted sums \Nmppc{} and \Npmt{} are calculated by integrating out the waveform for the main pulse (red).}
  \label{fig:pileup-unfolded}
  \vspace{-1\baselineskip}
\end{figure}

Figure~\ref{fig:pileup-unfolded} shows the unfolded pulses using the abovementioned techniques.
%The red waveform for the first unfolded pulse labelled as the ``main pulse'' in Figure~\ref{fig:pileup-unfolded} is determined as the waveform for the main photon because the preceding position and timing reconstructions are performed for the photon.
Once we have successfully unfolded the multiple pulses, we decide which pulses are eliminated based on the results of the preceding position and timing reconstructions\footnote{See Sections~6.4 and 6.5 of Ref.~\cite{meg2-detector}.}.
For the event shown above, the position and timing were reconstructed for the black-circled photon in Figure~\ref{fig:pileup-clustering}.
Therefore, the later pulse labelled the ``pileup pulse'' in Figure~\ref{fig:pileup-unfolded} is eliminated, and the first pulse (the ``main pulse'') remains for the subsequent energy reconstruction.
The weighted \Nmppc{} and \Npmt{} sums for the main pulse are calculated by integrating out the waveform of the main pulse.
%The multi-photon detection techniques are important to make the fitting robust as well as the energy resolution better.

%============================================
\subsection{Performance}
\label{subsec:pileup-performance}

\begin{figure}[tbp]
  \centering
  \includegraphics[width=8cm,clip]{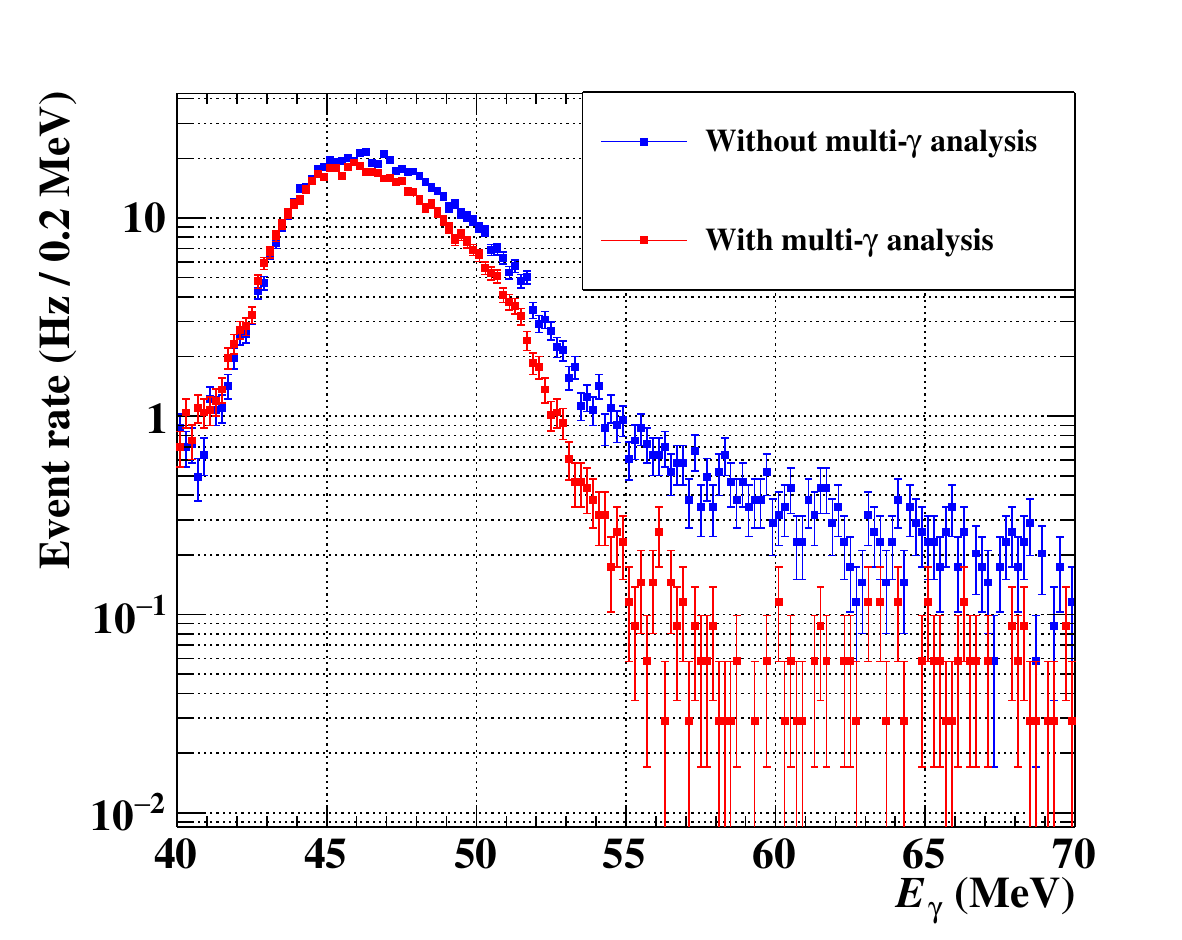}
  \caption{Background photon energy spectra with multi-photon elimination analysis (red) and without the analysis (blue) at a muon stopping rate of \qty{3.4e7}{\per\second}. An event drop below \qty{45}{\MeV} is derived from the trigger threshold based on photon energy.}
  \label{fig:bgspectrum}
  \vspace{-1\baselineskip}
\end{figure}

Successfully unfolded events are used for the physics analysis.
That is, events with coincident two photons and a fit failure are discarded from the analysis sample.
The performance of the multi-photon elimination is evaluated based on a reduction of background events in the analysis region (\(\egamma \in [\qty{48}{MeV}, \qty{58}{MeV}]\)) and the signal efficiency.

The number of background photons in \qtyrange[range-phrase=--,range-units=single]{48}{58}{\MeV} was reduced by \qty{34}{\percent}.
A clear drop around \qty{52.8}{\MeV} is seen in the energy spectrum thanks to the multi-photon elimination analysis (the red spectrum in Figure~\ref{fig:bgspectrum}).
The drop is essential to suppress the background events since the signal energy is monochromatic at \qty{52.8}{\MeV}.

The signal efficiency was evaluated as \qty{95}{\percent} based on the Monte Carlo simulation for the signal event.
The inefficiency of \qty{5}{\percent} is mainly due to events where photons in an electromagnetic shower escape from the shower and develop a new shower far enough from the original shower.
The new shower results in fake spatial on-timing peaks.

%=================================================
\section{Energy scale calibration}
\label{sec:egamma}

The energy scale of the calorimeter has to be calibrated so as not to miss signal events and to achieve a resolution as high as possible for the whole data-taking period.
The calibration for the 2021 data and the achieved energy resolution of \qty{1.8}{\percent} are described in Section~6.6 of Ref.~\cite{meg2-detector}.
%The calibration for the 2022 data is in progress.
%This paper focuses on a major difference between the 2021 and 2022 runs which was the Xe purity level and its temporal evolution during the run.
This paper focuses on xenon impurity and its temporal evolution during the 2022 run, which is an additional difficulty from the 2021 run.
This was due to the impure Xe which we added to fill the active volume up at the beginning.
Purification was done in parallel with the data-taking to remove impurities.

\begin{figure}[tbp]
  \centering
  \includegraphics[width=8cm,clip]{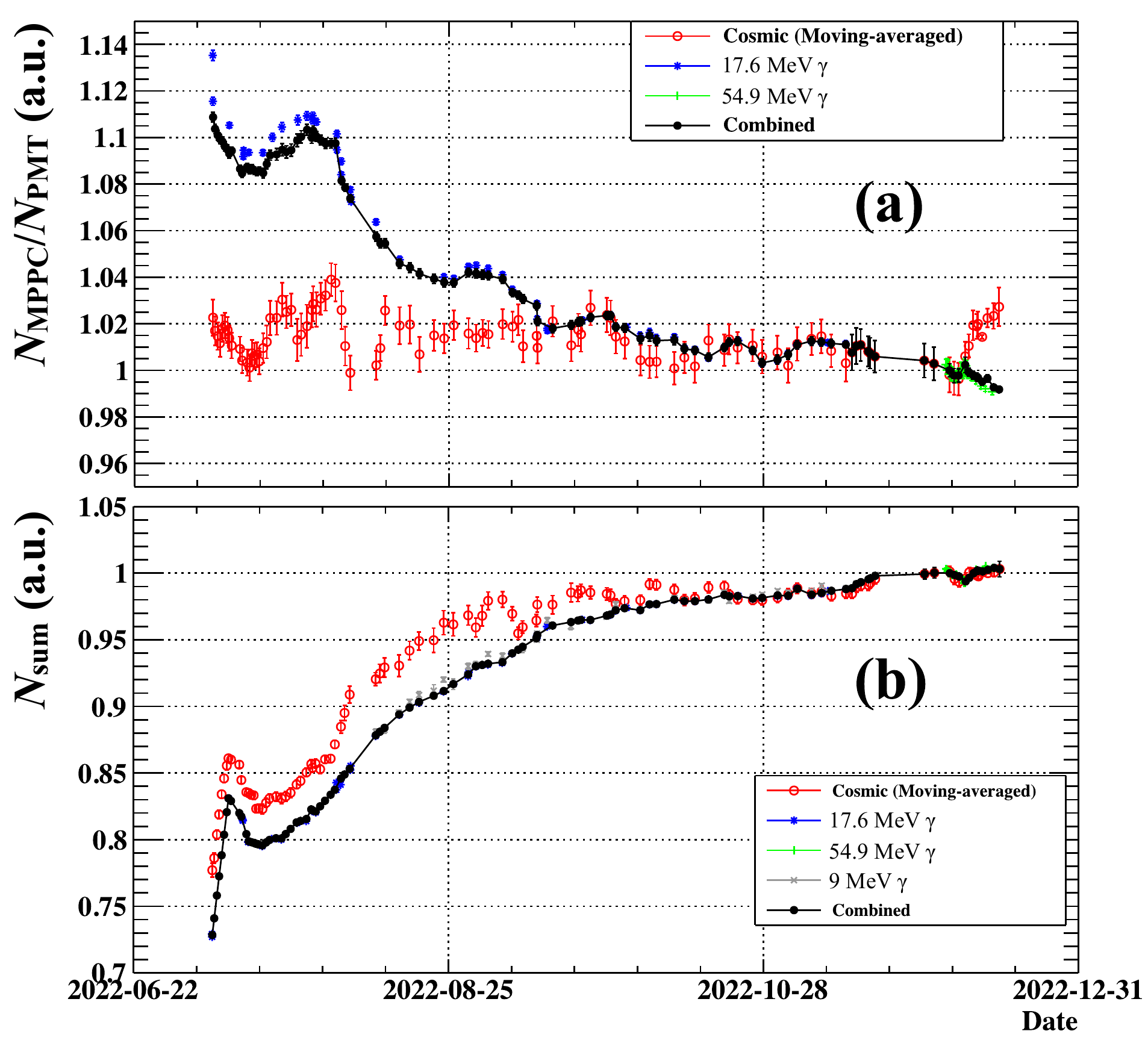}
  \caption{Temporal evolution of (a) a ratio of \Nmppc{} to \Npmt{} and (b) \Nsum{} during the 2022 run. The correction function \(r_{\mathrm{MPPC}}\) (\(T\)) is a reciprocal of the combined history of \(\Nmppc/\Npmt\) (\Nsum{}).}
  \label{fig:history}
  \vspace{-1\baselineskip}
\end{figure}

The impurity reduced the transparency of LXe to scintillation light.
It resulted in less scintillation light detection by PMTs for photons generated from the muon stopping target because most electromagnetic showers develop near the entrance face covered by MPPCs.
Position-dependent weights on \Npho{} mentioned in Section~\ref{sec:pileup}, however, were optimised to make the \Nsum{} peak width sharpest with \qty{54.9}{\MeV} photons from
%
%\begin{alignat*}{2}
\begin{equation*}
  \uppi^- \mathrm{p} \to \uppi^0 \mathrm{n},~\uppi^0 \to \upgamma \upgamma
\end{equation*}
%\end{alignat*}
%
at the last of the 2022 run.
We, therefore, need to calibrate the response difference between MPPCs and PMTs to keep the resolution using \qty{17.6}{\MeV} photons from \(^7\mathrm{Li}(\mathrm{p}, \upgamma)^8\mathrm{Be}\) reaction and cosmic rays penetrating the calorimeter as well as \qty{54.9}{\MeV} photons.
Note that the \qty{54.9}{MeV}-photon data was taken after the physics data-taking because it required a pion beam.
Figure~\ref{fig:history}(a) shows a temporal evolution of the ratio of \Nmppc{} to \Npmt{} during the 2022 run.
The ratio for photon data decreased as the Xe purity was recovered, as discussed above.
Meanwhile, the ratio in cosmic-ray data was constant during the run because scintillation light was also generated near PMTs.
%\footnote{The scintillation light is emitted along a cosmic-ray path.}.
The correction function $r_{\mathrm{MPPC}}$ is a reciprocal of the combined \Nmppc{}/\Npmt{} history drawn in black in Figure~\ref{fig:history}(a).

The total detected scintillation light also decreased as impurities in LXe increased.
It was calibrated also using \qty{9}{\MeV} photons from \(^{58}\text{Ni}(\mathrm{n}, \upgamma)^{59}\text{Ni}\) reaction as well as \qty{17.6}{\MeV} and \qty{54.9}{\MeV} photons and cosmic rays.
Figure~\ref{fig:history}(b) shows the temporal evolution of the \Nsum{} peak positions.
We see a good agreement in the temporal evolutions between \qty{9}{\MeV} and \qty{17.6}{\MeV} photons.
On the other hand, there is a discrepancy between photons and cosmic rays, which is not understood.
We, however, relied on the photon data since the shower development should be similar to that of signal and background photons in a muon beam.
The cosmic-ray data was used to connect the peaks with \qty{17.6}{\MeV} and \qty{54.9}{\MeV} photons.
The stability of the energy scale during the physics data-taking was assessed to be \qty{0.2}{\percent} based on the standard deviation of reconstructed energy for \qty{17.6}{\MeV} photons after applying the correction function \(T\).

%=================================================
\section{Conclusion}
\label{sec:conclusion}

The MEG~II LXe calorimeter with \qty{900}{\liter} of LXe and \num{4760} VUV-sensitive photosensors measures photon position, timing, and energy.
This paper focuses on photon energy reconstruction, discussing the multi-photon elimination algorithm and energy scale calibration.

The multi-photon elimination algorithm is performed to reconstruct single-photon energy in a high-intensity muon beam.
Pileup photons are unfolded by the template waveform fit with three multi-photon detection techniques.
The number of background events in the range of [\qty{48}{\MeV}, \qty{58}{\MeV}] was reduced by \qty{34}{\percent} in data while keeping the signal efficiency as high as \qty{95}{\percent}.

We discuss the temporal evolution of the response difference between MPPCs and PMTs and of the energy scale due to Xe impurity.
The stability during the physics data-taking was evaluated as \qty{0.2}{\percent}.
The calorimeter performance, such as the resolution, will be evaluated after the ongoing calibration.
The physics analysis results with the 2021 and 2022 data will finally be released this year.

%=================================================
\section*{Acknowledgements}
\label{sec:acknowledgement}

We are grateful for the support provided by PSI as the host laboratory.
This work was supported by
MEXT/JSPS KAKENHI Grant Numbers 21H04991, 22K21350, 24KJ0607, and
JSPS Core-to-Core Program, A. Advanced Research Networks JPJSCCA20180004.

%=================================================
% BibTeX or Biber users please use (the style is already called in the class, ensure that the "woc.bst" style is in your local directory)
\bibliography{mybibfile} % Replace "your_bib_file" with the actual name of your .bib file

\begin{thebibliography}{6}

\bibitem{Kuno-okada}
Y.~Kuno, Y.~Okada, Muon decay and physics beyond the standard model, Reviews of Modern Physics \textbf{73}, 151 (2001). \doiwoc{10.1103/RevModPhys.73.151}

\bibitem{meg2-design}
A.M. Baldini et~al., {The design of the MEG~II experiment}, Eur. Phys. J. C \textbf{78}, 380 (2018). \doiwoc{10.1140/epjc/s10052-018-5845-6}

\bibitem{meg2-detector}
K.~Afanaciev et~al., {Operation and performance of the MEG~II detector}, Eur. Phys. J. C \textbf{84}, 190 (2024). \doiwoc{10.1140/epjc/s10052-024-12415-3}

\bibitem{meg2-2021}
K.~Afanaciev et~al., {A search for \(\upmu^+ \to \upe^+ \upgamma\) with the first dataset of the MEG~II experiment}, Eur. Phys. J. C \textbf{84}, 214 (2024). \doiwoc{10.1140/epjc/s10052-024-12416-2}

\bibitem{vuv-mppc}
K.~Ieki et~al., {Large-area MPPC with enhanced VUV sensitivity for liquid xenon scintillation detector}, Nucl. Instrum. Meth. A \textbf{925}, 148 (2019). \doiwoc{10.1016/j.nima.2019.02.010}

\bibitem{drs4}
S.~Ritt, {The DRS chip: Cheap waveform digitizing in the GHz range}, Nucl. Instrum. Meth. A \textbf{518} (2004). \doiwoc{10.1016/j.nima.2003.11.059}

\end{thebibliography}
%
% Non-BibTeX users please use
%
%\begin{thebibliography}{}
%
% and use \bibitem to create references.
%
%\bibitem{RefJ}
% Format for Journal Reference
%Journal Author, Article title. Journal \textbf{Volume}, page numbers (year). \url{https://doi.org/Article-DOI-number}
% Format for books
%\bibitem{RefB}
%Book Author, \textit{Book title} (Publisher, place, year) page numbers
% etc
%\end{thebibliography}

\end{document}